\documentclass[final,3p,times,twocolumn]{elsarticle} 

\usepackage{amsmath}  
\usepackage{amsfonts} 
\usepackage{graphicx} 
\usepackage{siunitx}
\usepackage[utf8]{inputenc}
\usepackage{subcaption}
\usepackage[T1]{fontenc}
\usepackage{placeins} 
\usepackage{balance}

\newcommand{\um}{\si{\micro\meter}}

\journal{Radiation Measurements}

\usepackage{wasysym}
\usepackage{booktabs}

\begin{document}

\sloppy

\title{Pixelated silicon detector for radiation beam profile measurements}

\author[1,2,3]{J.~Tikkanen\corref{cor1}}
\ead{joonas.tikkanen@stuk.fi}
\cortext[cor1]{Corresponding author}
\author[2,3]{S.~Kirschenmann}
\author[2,4]{N.~Kramarenko}
\author[2,3]{E.~Brücken}
\author[2]{P.~Koponen}
\author[2,4]{P.~Luukka}
\author[1,2,3]{T.~Siiskonen}
\author[2]{R.~Turpeinen}
\author[2,5]{J.~Ott}

\address[1]{Radiation and Nuclear Safety Authority (STUK), Jokiniemenkuja 1, FI-01370 Vantaa, Finland}
\address[2]{Helsinki Institute of Physics, Gustaf Hällströmin katu 2, FI-00014 University of Helsinki, Finland}
\address[3]{University of Helsinki, Department of Physics, P.O. Box 64, FI-00014 University of Helsinki, Finland}
\address[4]{Lappeenranta-Lahti University of Technology LUT, Yliopistonkatu 34, FI-53850 Lappeenranta, Finland}
\address[5]{Santa Cruz Institute for Particle Physics, University of California Santa Cruz, 1156 High Street, CA-95064, USA}

\begin{abstract}

A pixelated silicon detector, developed originally for particle physics experiments, was used for a beam profile measurement of a Co-60 irradiator in a water phantom. The beam profile was compared to a profile measured with a pinpoint ionization chamber. The differences in the pixel detector and pinpoint chamber relative profiles were within approximately 0.02, and after calculating correction factors with Monte Carlo simulations for the pixel detector, the differences were decreased to almost less than 0.005. The detector's capability to measure pulse-height was used to record an electron pulse-height spectrum in water in the Co-60 beam, and the results agreed well with simulations.

\end{abstract}


\maketitle

\section{Introduction}

An accurate estimation of dose delivered to the patient in radiation therapy requires measurement of dose-distribution, or beam profile, in a water phantom. Whereas determination of the dose at the center of the beam for field sizes of for example 10~cm$\times$10~cm can be done with traditional farmer type ionization chambers with nominal volume of 0.6~cm$^3$, small and conformal beams used in modern radiotherapy require a small detector to minimize the averaging of dose in the detection volume \cite{TRS483}. Therefore, the quality assurance requirements for the modern radiotherapy beams are challenging to fulfill with traditional measurements based on ionization chambers. Small pinpoint chambers or diamond detectors may be used \cite{TRS483}, but the measurements can be cumbersome. Position sensitive detectors with a good position resolution and well-known energy and dose rate dependence would be a valuable addition to rapidly scan the beam to obtain dose profiles.

\begin{figure*}[ht]
    \centering
    \includegraphics[width=0.8\textwidth]{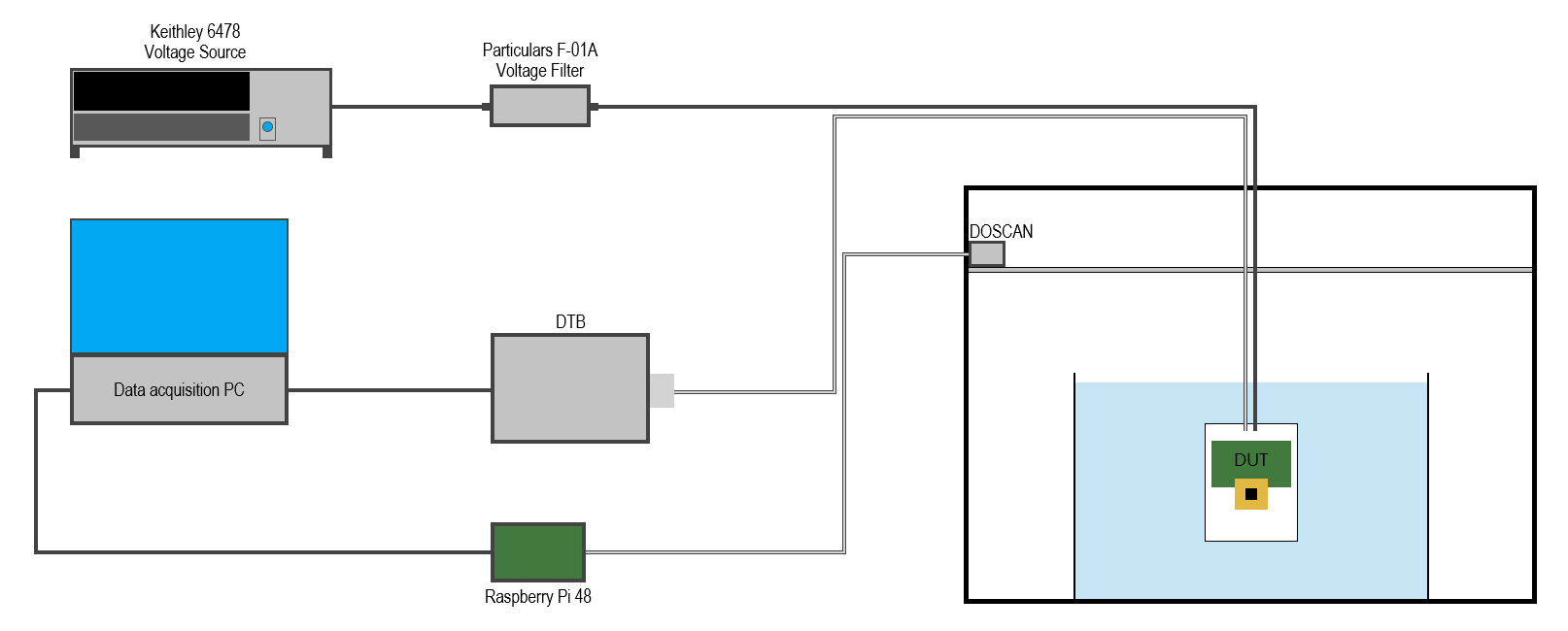}
    \caption{Schematic image of the data acquisition chain. Power cables are not depicted.}
    \label{schematic}
\end{figure*}

An alternative to moving a detector with a single radiation detecting volume is to use detectors divided into segments, either strips or pixels. Pixelated detectors can provide an excellent two-dimensional spatial resolution, while a sub-millimeter pixel size helps to reduce the individual pixel count rates that can be very high in radiotherapy beams. Acquisition of the energy spectrum would possibly give additional information about possible changes in the beam quality.

Pixelated semiconductor detectors have been used as tools in particle physics, for example at the CMS experiment at CERN for tracking charged particles. In tracking sensors collision products ideally do not experience significant energy loss, hence the material of choice should be light. In radiation therapy, when measuring dose-to-water, the detector should not affect the particle fluence significantly compared to only water being present, and hence similar considerations apply \cite{Attix}. The atomic composition and density of silicon are not as close to water as those of ionization chambers, but due to a high amount of ion-hole pairs created per energy deposited, the detector can be kept thin to minimize the effect on the particle fluence. Also, silicon as a semiconductor material allows fast signal processing, and acquisition of a pulse-height spectrum. 

In this study, we used a silicon pixel detector, to measure the beam profile of a Co-60 irradiator. This detector-type was originally fabricated for the R\&D phase of the first upgrade of the inner tracking detector of the CMS experiment \cite{CMSTrackerGroup:2020edz}. Co-60 irradiators were common in radiation therapy before the development of linear accelerators, and nowadays, they are used mostly for detector calibration purposes. The energies of photons emitted by a Co-60 source (1.173~MeV and 1.332~MeV) are close to the mean energy of a 6~MV linear accelerator beam \cite{Chow2016}. The beam profile measured with the pixel detector was compared to a profile measured with a pinpoint ionization chamber. Also, we measured a 2D-profile to demonstrate the capability of the pixelated detector for high resolution planar profile measurements. The pulse-height measurement feature of the detector was used to record a pulse-height spectrum of the detector in water in the Co-60 beam, and the results were compared to a simulated spectrum.

\section{Measurement setup and methods}

In the beam profile measurements, we used a GBX200 (Best Theratronics Ltd, Ottawa, Canada) Co-60 irradiator at the premises of the Finnish Radiation and Nuclear Safety Authority (STUK) dosimetry laboratory. The irradiator contained a high-activity Co-60 source, and the radiation beam was collimated with inter-meshing lead and tungsten bars. The radiation beam shape was rectangular, beam axis horizontal, and the nominal field size 10~cm$~\times$~10~cm at 100~cm source-to-detector distance.

The dose profiles were measured in a 30~cm$~\times$~30~cm$~\times$~30~cm water phantom with a 1~cm thick front wall. The phantom wall material was poly-methyl-methacrylate (acrylic glass, PMMA), and because of the horizontal beam direction, the beam passed through the front wall before reaching water. Distance from the source to the phantom wall (source to surface distance, SSD) was 95~cm. The water temperature during the measurements was $19\pm1~^\circ$C. 

The coordinate system in the rest of this paper is the following: horizontal (x) axis is the horizontal axis perpendicular to the beam axis, and the depth (z) axis is the beam axis. The y-axis is vertical. To move detectors inside the water phantom, a scanning device capable of moving the detector in three dimensions was built from a 3D-printer (Creality Ender 5 plus, Shenzhen, China), named DOSCAN \cite{scanner}. The scanner used stepper motors, and self-made control electronics.

\subsection{Silicon pixel detector}
\begin{figure*}[ht]
    \centering
    \includegraphics[height=6.8cm]{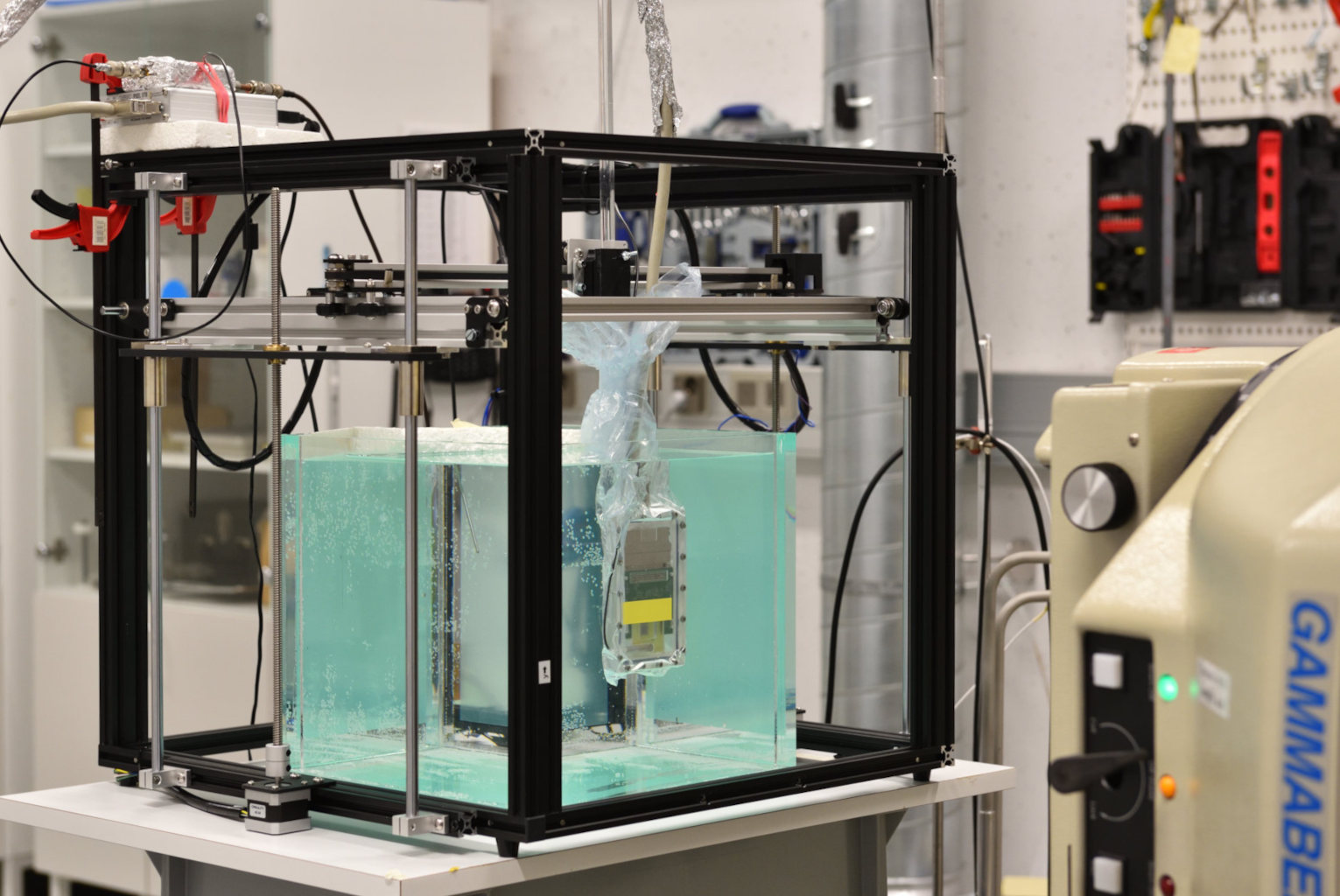}
    \includegraphics[height=6.8cm]{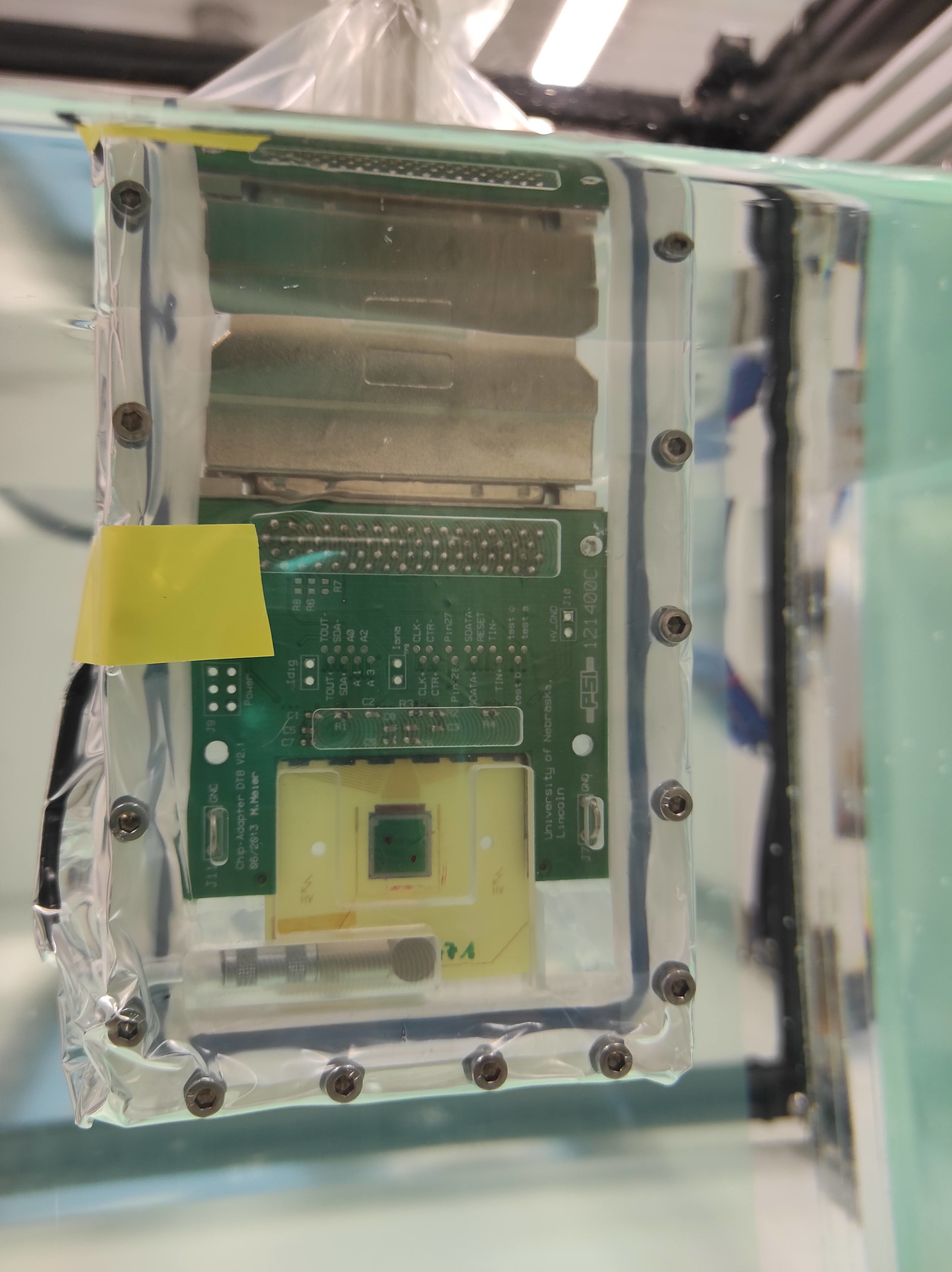}
    \caption{Measurement setup used for the beam profile scans. The pixel detector is mounted inside the acrylic box. The detector box is attached to the 3D scanning system surrounding the phantom via a PMMA rod. On the right, in addition to the detector chip, printed circuit board (PCB, yellow), horseshoe connector (green), and the LEMO cable for detector bias voltage can be seen. On the left, partly visible are the GBX200 Co-60 irradiator and the readout electronics on top of the scanner.}
    \label{measurementsetup}
\end{figure*}
To fulfill the requirements for the performance of the detector, a 280~\um\ thick n$^+$ in n type silicon-based pixel sensor design with a rectangular pitch of $100\times150$~\um$^2$ and $52\times80$ pixels, with a size of 1$\times$1~cm$^2$, and active area of 0.78$\times$0.8~cm$^2$ was chosen. The detector is described in more detail in \cite{CMSTrackerGroup:2020edz}. Si pixel detectors of this design are used in CMS tracker at CERN, which consists in total of 66560 pixels organized into 16 sections of $52\times 80=4160$ pixels each. The sensor used for the beam profile measurements forms a Single Pixel Chip Module (SCM) when bonded with a single readout chip (ROC), using indium solder bumps.

The ROC (psi46dig2.1-r \cite{Kastli2006}) contains an individual readout for each pixel, called a pixel unit cell (PUC). The pulse-height of charge created by a charged particle is measured in each PUC. The pulse processing chain includes pre-amplification, pulse shaping circuits and a charge discriminator, and the measured pulse-height is converted into a digital signal (8-bit). The readout from the PUCs is done in double columns (2$\times$80 pixels). If a pixel has measured a signal (has been hit), it sends the pulse-height information to the double-column readout, from where it is sent further into a readout buffer. The data is written finally into event frames, where one event contains the pixel addresses (column and row) and the pulse-heights from all pixels hit in a 25~ns time window.

The full electronic readout chain in the measurements (shown in Fig.~\ref{schematic}) consisted of an adapter card that connects the pixel detector via standard SCSI cable to the so-called Detector-Test-Board (DTB) based on an Altera FPGA \cite{Spannagel:2162902}. The DTB was connected via a USB2 cable (15\,m long) to the data acquisition PC.
The bias was applied from the back side of the detector with a Keithley 6478 Picoammeter/Voltage source via a high voltage filter (Particulars F-01A) close to the detector. The data acquisition PC as well as the voltage source were located in a separate room to control the setup remotely during measurements when the experimental area is closed due to radiation protection.

To characterize and train the pixel detector before the measurement, certain tests are undertaken, which are described extensively in \cite{Spannagel:2137512}. These tests ensure the functioning state of the ROC, calibrate the pixels and deliver additional information, such as the trimming value, which is set above the noise level to prevent false detection. In the measurements, the trimming value was set to 60 VCAL, which corresponds to low energy threshold of approximately 10~keV. Unfortunately, the ROC does not allow self-triggering. Instead, trigger signals are sent at a constant rate of 100~kHz, and only the latest event is read out with each trigger.

\subsection{Measurements with the pixel detector}

For the beam profile measurements we used the setup shown in Fig. \ref{measurementsetup}. The silicon pixel detector was inserted in a custom designed and watertight CNC-machined box of acrylic glass (PMMA). The box was shaped to minimize the amount of air around the detector chip while maintaining enough space to have a low probability of breaking the detector during setup, and there was 5~mm of air in front, and 1~mm behind the detector. The box was attached via a PMMA rod to the DOSCAN scanner. Two thin plastic bags were placed around the casing, the rod, and cables to ensure water-tightness of the system. The detector shorter edge (7.9~mm) was along the horizontal axis.

The detector was still during, and moved between single measurements. The detector was at 5~cm distance (depth) from the phantom front wall, and the step size of the detector 4~mm, approximately half of the silicon chip width, leading to overlap of the detector between measurement points. The depth of the detector in the phantom was measured with a Micro-epsilon (Ortenburg, Germany) IDL1700 laser displacement sensor, with a knowledge of the distance of the detector chip from the front wall of the casing. The stepper motors were turned off during the measurements, as described in \cite{scanner}, due to noise induced in the detector.

A 2D scan of the beam profile was made of the upper left corner of the beam at 5~cm depth. The scan area was approximately -8~cm to center on the x-axis and from center to 8~cm on the y-axis, the step size being 8~mm on both axes. The horizontal axis was scanned first before taking a step in the vertical direction, leading to minimal movement of the less-ideally functioning vertical axis \cite{scanner}.

A laptop handled data acquisition through the pXar software \cite{Spannagel:2137512}, and the scanner was operated with a Raspberry~Pi~4B (Raspberry Pi Foundation, Cambridge, UK). The laptop gave a signal to the Raspberry, through a socket connection when the data acquisition finished, and the Raspberry to the laptop after the detector had been moved to the correct position. The code on the laptop was written in C++, utilising pXar and Root libraries \cite{Brun1997}, and the code on Raspberry with python.

\subsection{Measurements with an ionization chamber}

The GBX200 beam profile was also measured with an ionization chamber and the DOSCAN scanner. The chamber employed was a PTW 31015 pinpoint (Freiburg, Germany), with cavity radius of 1.45~mm and length of 5~mm. The chamber current was measured with a Keithley 6517B electrometer (Tektronix, Beaverton, USA). The chamber reference point was placed at 5~cm depth from the phantom wall and the current was measured at 1~mm intervals on the x-axis. The field size and phantom orientation were the same as in the pixel detector measurements.

In a 2D scan of the corner of the beam with the pinpoint chamber, the step sizes were 2~mm on the horizontal, and 4~mm on the vertical axis. The detector was in the beginning of the scan placed in the center of the beam, indicated by the cross-hair of the GBX200. The scanning order was, again, horizontal axis first.

\subsection{Simulations}

The silicon pixel detector inside the watertight casing was modelled with EGSnrc radiation transport simulation software \cite{egsnrc}, using the C++ class library and egs\_chamber user code \cite{Wulff2008a}. A model of the GBX200 irradiator, constructed with BEAMnrc user code \cite{beamnrc}, was used as a shared library source in the simulations. The modeling of the GBX200 irradiator will be described in a further study.

The model of the detector included the silicon chip, gap for the bump-bonds, ROC, PCB, horseshoe-connector, and the casing. The connector and cables were not modeled, and those areas were filled with PMMA. These areas had more than 5~mm of plastic in front of the detector, and although the simplification affects the photon flux slightly, the electrons from these regions do not reach the silicon chip. The dose was simulated in a 0.9~mm~$\times$~6~mm~$\times$~0.28~mm region, the size of $6\times60$ pixels, at the center of the chip. The detector was shifted in 2~mm intervals on the x-axis from the center of the beam to 78~mm. 

The variance reduction techniques (VRT) for the egs\_chamber user code are described in \cite{Wulff2008a}. An intermediate phase space file was scored to a volume that enveloped at least 1.5~cm of the plastic in the casing around the hole cut for the detector. The cross-section enhancement VRT was used in a 2.5~cm sized cube around the detector chip with an enhancement factor of 64. Russian roulette VRT was also used, with air around the detector as cavity geometry, rejection factor of 128, water as rejection range medium, and Esave parameter of 521~keV. Electron and photon cut-offs (ECUT and PCUT) were 521~keV and 10~keV.

The simulation was also run including only the detector chip, ROC and PCB in the water phantom to estimate the performance of the detector with an ideal (water equivalent) casing. Also, a dose-to-water profile was simulated in 0.2~cm~$\times$~0.6~cm~$\times$~0.1~cm (xyz) sized water voxels along the x-axis.

A pulse-height spectrum was simulated using the egs\_phd user code. The active volume was the size of the detector chip excluding the outer rim of pixels. The bin-size was 10~keV, the electron cut-off parameter 0.531~MeV, and photon cut-off energy 10~keV. Monte Carlo transport parameters common to all the simulations are given in Table \ref{MCparam}.

\begin{table}[!ht]
	\caption{\label{MCparam} Monte Carlo transport parameters for the EGSnrc simulations.}
	\centering
		\begin{tabular}{@{}ll}
		\toprule
	Global SMAX & 1e10                    \\
	ESTEPE & 0.25                         \\
	XImax & 0.5                           \\
	Skin depth for BCA & 3                \\
	Boundary crossing algorithm & EXACT   \\
	Electron-step algorithm & PRESTA-II   \\
	Spin effects & On                     \\
	Brems angular sampling & KM           \\
	Brems cross sections & NIST           \\
	Photon cross sections & mcdf-xcom     \\
	Electron Impact Ionization & On       \\
	Triplet production & On               \\
	Radiative Compton corrections & On    \\
	Bound Compton scattering & On         \\
	Pair angular sampling & KM            \\
	Pair cross sections & NRC             \\
	Photoelectron angular sampling & On   \\
	Rayleigh scattering & On              \\
	Atomic relaxations & On               \\
	Photonuclear attenuation & On         \\
		\bottomrule
		\end{tabular}
\end{table}

\begin{figure}[ht]
    \centering
    \includegraphics[width=0.46\textwidth]{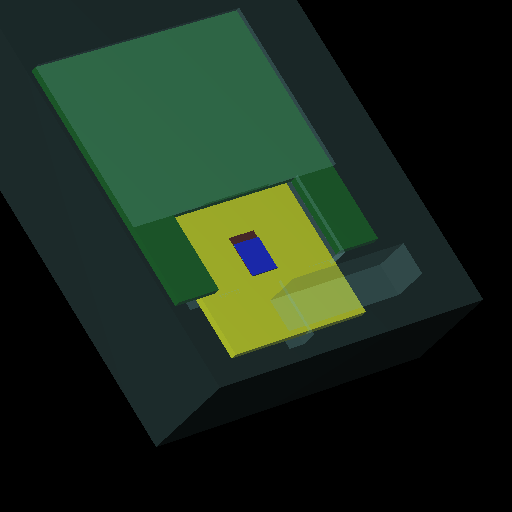}
    \caption{Visualisation of the Monte Carlo model of the detector and the waterproof casing.}
    \label{Simulation_model}
\end{figure}

\section{Results and discussion}

\subsection{Comparison of measured dose-profiles}

\begin{figure*}[!ht]
    \centering
    \includegraphics[width=0.9\textwidth]{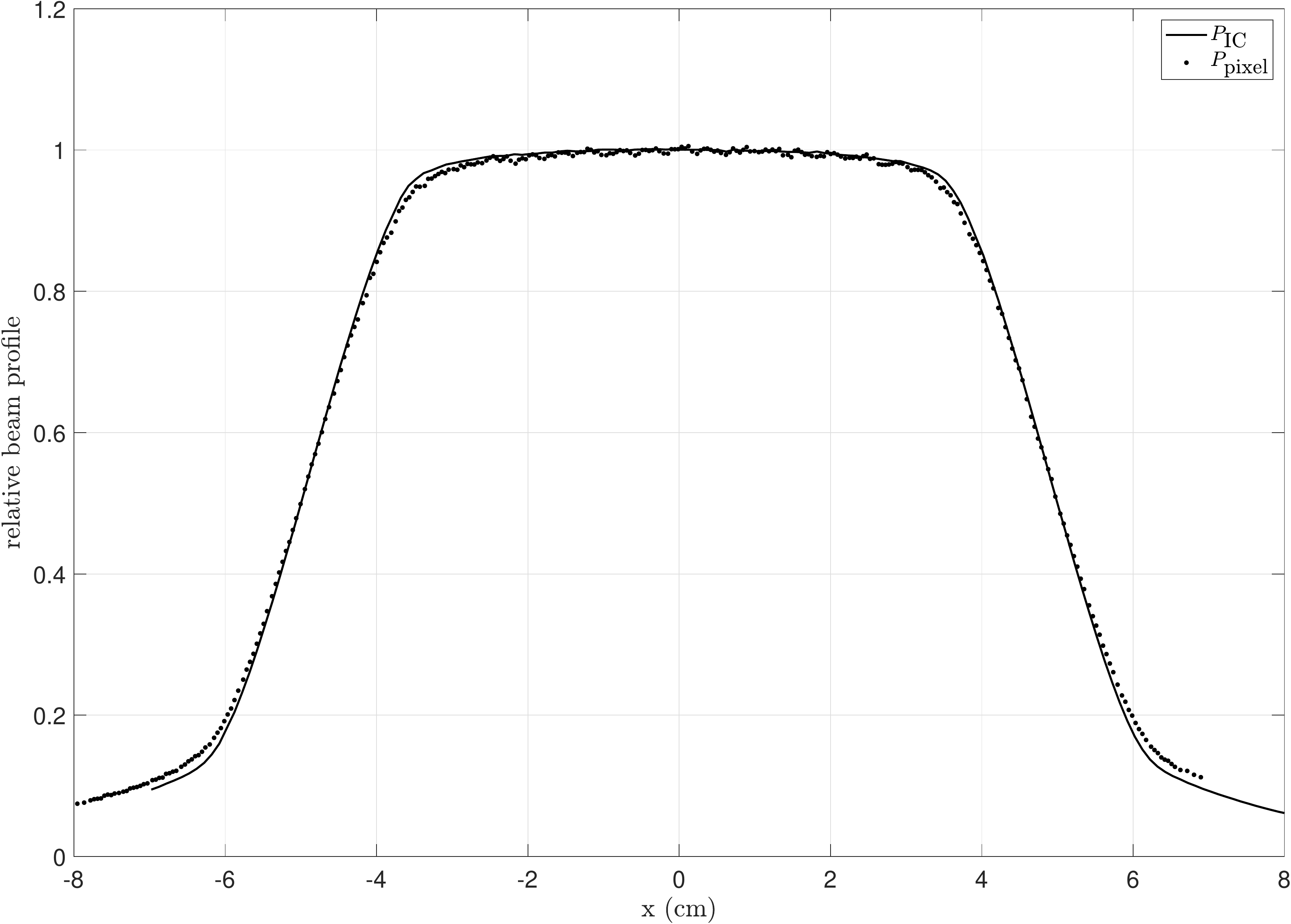}\\
    \hspace{-0.42cm}    
    \includegraphics[width=0.9182\textwidth]{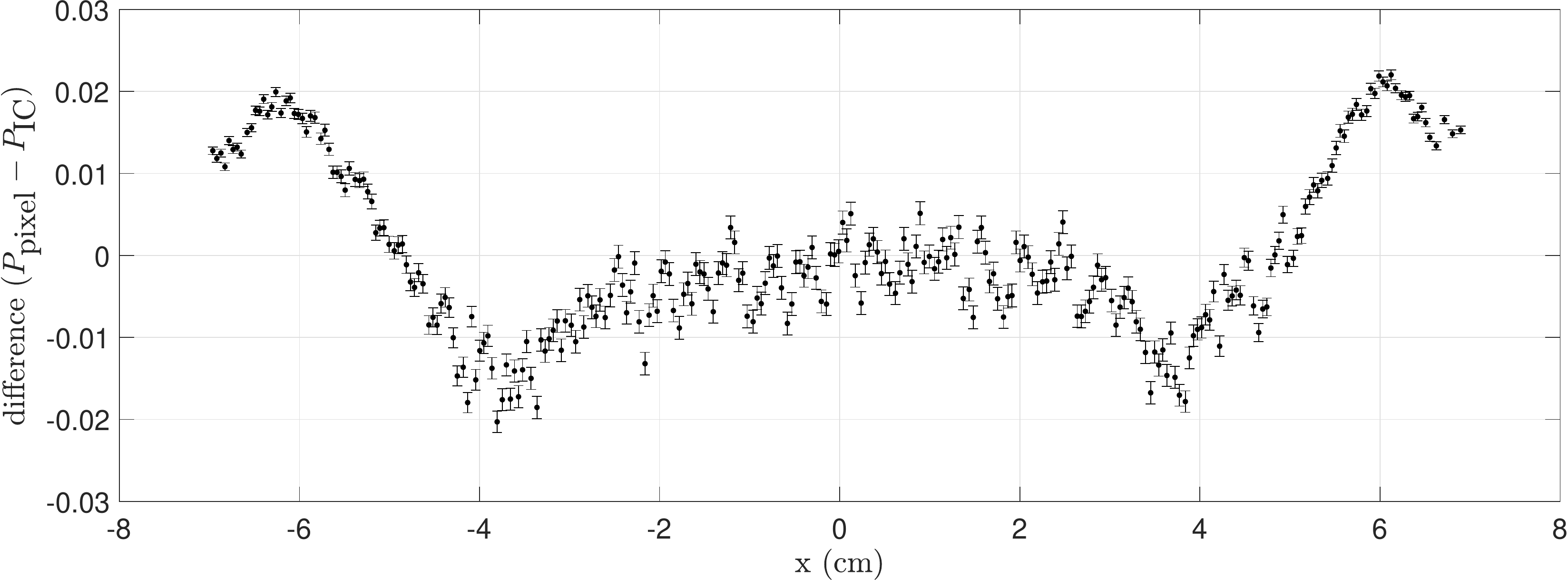}
    \caption{The relative current-profile of the pinpoint ionization chamber ($P_\textrm{IC}$), and hit-profile of the pixel detector ($P_\textrm{pixel}$) in the GBX200 beam, at 5 cm depth in a water phantom, and difference between the two profiles. The uncertainty ($k=1$) was calculated assuming the chamber measurement uncertainty to be insignificant, and the pixel detector uncertainty was calculated from Poisson statistics.}
    \label{profiles}
\end{figure*}

The relative beam profiles measured with the ionization chamber and the pixel detector are shown in Fig. \ref{profiles}. The relative profiles are denoted as $P_\textrm{IC}(x)=I_\textrm{IC}(x)/I_\textrm{IC}(0)$ and $P_\textrm{pixel}(x)=I_\textrm{pixel}(x)/I_\textrm{pixel}(0)$, where $I_\textrm{IC}(x)$ is the current of the chamber, and $I_\textrm{pixel}(x)$ the number of recorded events per pixel (hit-profile) at position $x$. The values for the pixel detector are averaged over six columns (area of 0.9~mm $\times$ 8~mm). For both profiles, the normalization value ($I_\textrm{IC}(0)$ and $I_\textrm{pixel}(0)$) was an average from the profile around the center of the beam. The uncertainty of the silicon detector profile was calculated assuming Poisson statistics, meaning that the standard uncertainty is the square-root of the number of hits. However, the standard deviation of the hits in the pixels was larger, as can be seen in Fig. \ref{profiles}. Even though the ionization chamber current profile should give a good approximation of the dose-profile, charged particle fluence effects and dose averaging in the chamber cavity volume affect the response of the detector at the beam edges, and how much the chamber current profile differs from the dose-profile is unknown to us.

The field sizes (full width at half maximum, FWHM) were calculated by finding the half-value at both edges of the beam from a second order polynomial to the data around the half-value. The FWHM values for the profiles in Fig. \ref{profiles} were 10.00~cm for the chamber and 10.01~cm for the pixel detector. The x-axis values were shifted with the half-values so that the zero position was at half-way between the half-value positions. This allowed the comparison of different profile measurements without the effect of detector positioning.

The pixel detector measurements were in a non-ideal geometry due to the air inside the water-tight casing. Although the casing material is relatively water equivalent (PMMA), there was roughly 5~mm of air in front, and 1~mm behind the detector. The air around the detector affects the electron and photon flux compared to only water being present, and the effective depth of the detector was not 5~cm. 

From the simulations, we calculated two different correction factors for the measured data: the ratio between dose-to-water and dose to detector relative profiles
\begin{equation}
    k_1(x)=\frac{P_{D_\textrm{w}}(x)}{P_{D_\textrm{Si}}(x)} = \frac{ D_\textrm{w}(x)/D_\textrm{w}(0) }{ D_\textrm{Si}(x)/D_\textrm{Si}(0) },
\end{equation}
and between dose-to-detector without and with the casing
\begin{align}
    k_2(x)=&\frac{P_{D_\textrm{Si, no casing}}(x)}{P_{D_\textrm{Si}}(x)} \nonumber \\ =& \frac{ D_\textrm{Si, no casing}(x)/D_\textrm{Si, no casing}(0) }{ D_\textrm{Si}(x)/D_\textrm{Si}(0) },
\end{align} 
where $D$ refers to simulated dose. The factor $k_1$ is akin to a calibration factor for the pixel detector. The second one is for speculation about how the detector would perform in a more ideal geometry, without air, or other non water-equivalent materials around the detector.

\begin{figure*}[!ht]
    \centering
    \includegraphics[width=0.8\textwidth]{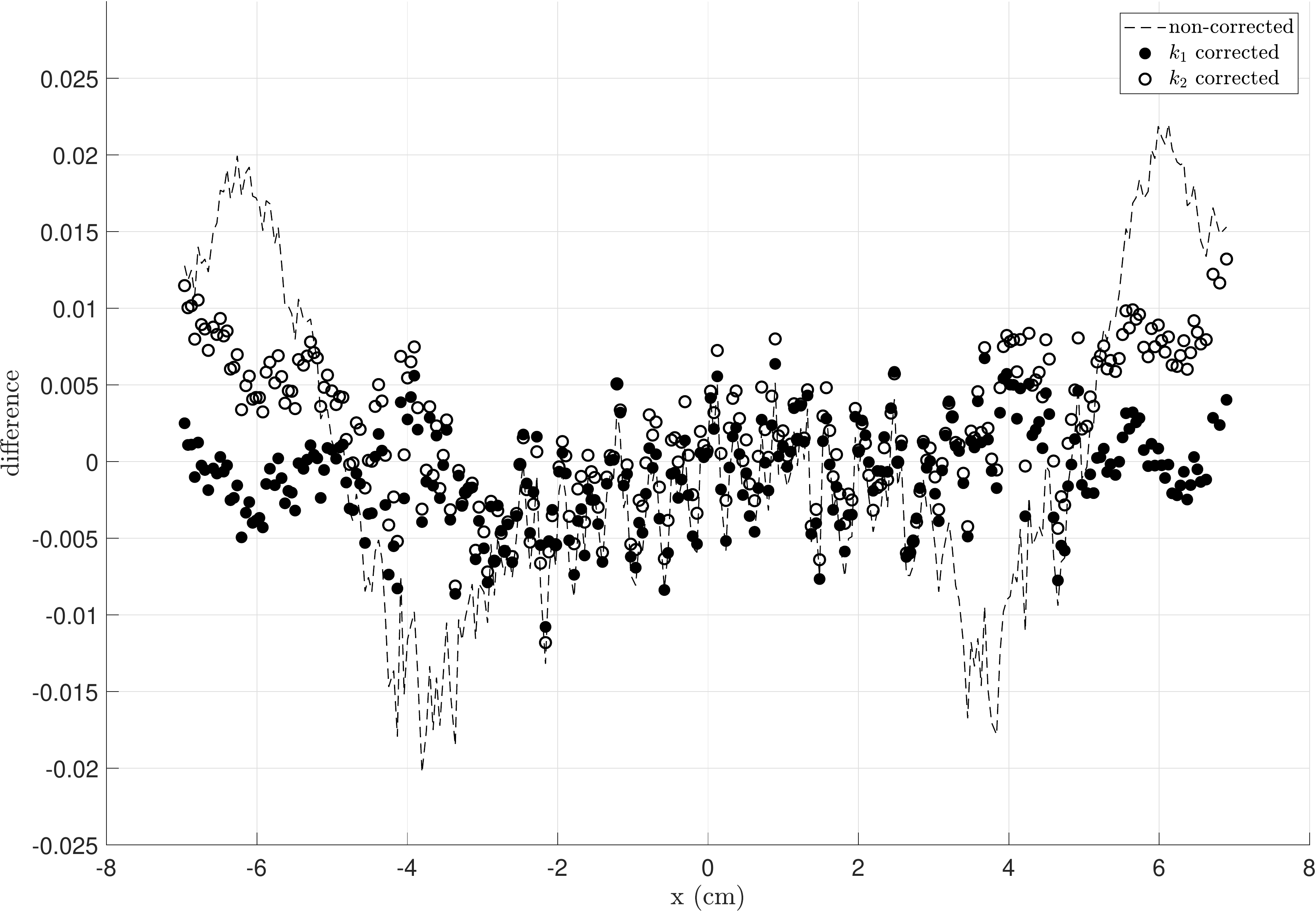}
    \caption{The difference of the pixel detector beam profile, non-corrected and multiplied with the factors $k_1$ and $k_2$ obtained from simulations, to the pinpoint ionization chamber beam profile. Error bars are omitted for clarity, but as in Fig. \ref{profiles}, the standard deviation of the data was significantly larger than the statistical uncertainty.}
    \label{profiles_corrected}
\end{figure*}

The difference of the relative silicon detector profile $P_\textrm{pixel}$ multiplied with the factors $k_1$ and $k_2$ to $P_\textrm{IC}$ is depicted in Fig. \ref{profiles_corrected}. For the $k_1$ corrected profile, the difference stays almost below 0.01 and for the most part is between $\pm$0.005. This means that the deviation between the non-corrected pixel detector and chamber profile is likely due to detector response, and not due to errors in the measurements. In addition, it means that the detector is capable of measuring profiles with sufficient accuracy if a calibration is given as a function of position, at least in a Co-60 beam.

The $k_2$ corrected $P_\textrm{pixel}$ profile is also significantly closer to $P_\textrm{IC}$. This means most likely that the air around the detector affected the profile measurement, and with a more optimal casing, the dose-profile could be measured with a higher accuracy without applying position dependent correction factors.

Using the charge deposited in a pixel at each detector location (charge-profile) instead of number of detection events (hit-profile) should give a better estimation of the dose to the detector chip since this value is approximately proportional to the energy imparted in the pixel. However, both methods yielded similar results, except the charge-profile had a higher variance. Therefore the hit-profile was used.

\begin{figure*}[!ht] % Colored figures cost extra?
    \centering
     \begin{subfigure}[c]{0.45\textwidth}
        \centering
        \includegraphics[width=\textwidth]{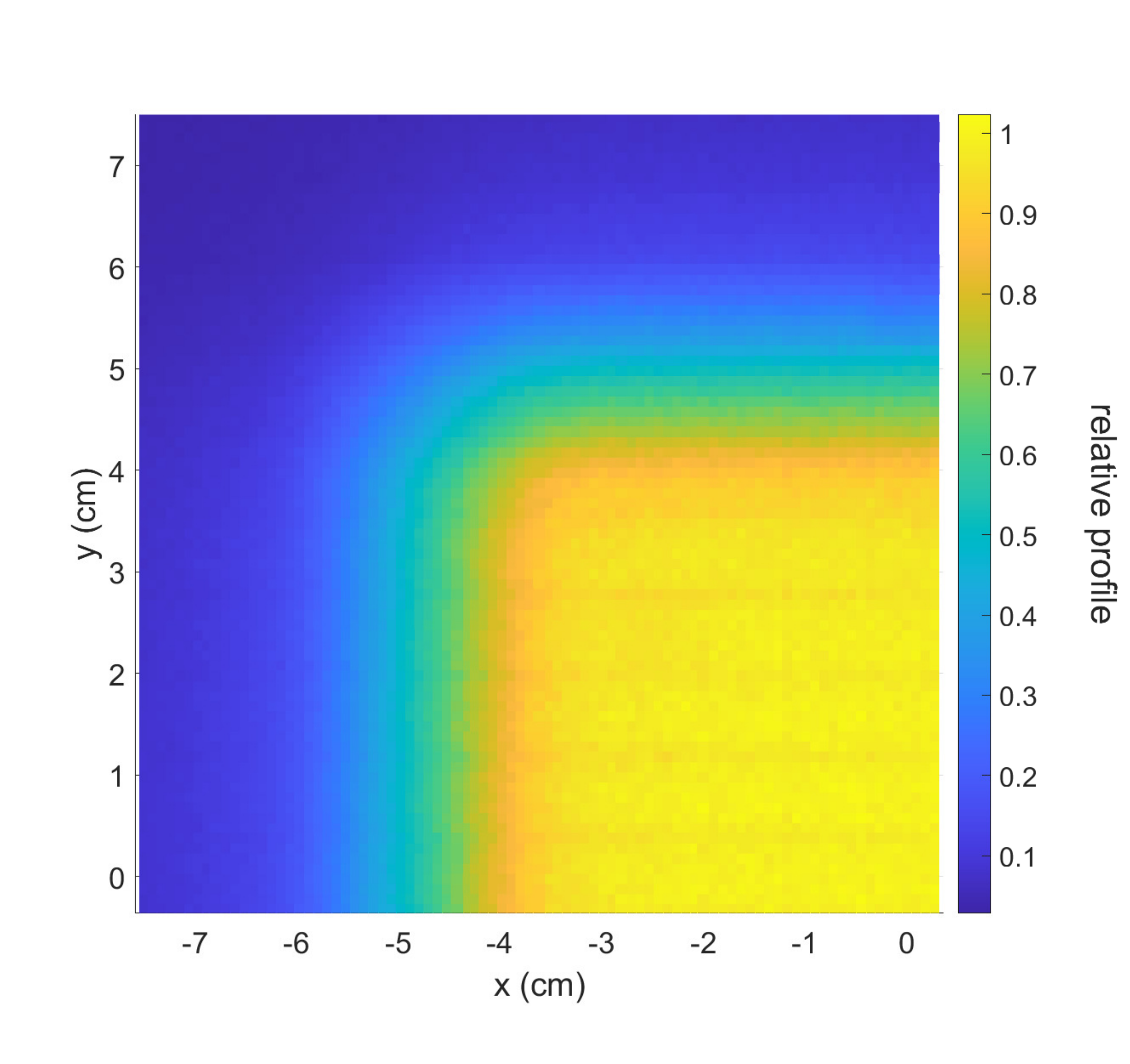}
        \caption{}
    \end{subfigure}
    \begin{subfigure}[c]{0.45\textwidth}
        \centering
        \includegraphics[width=\textwidth]{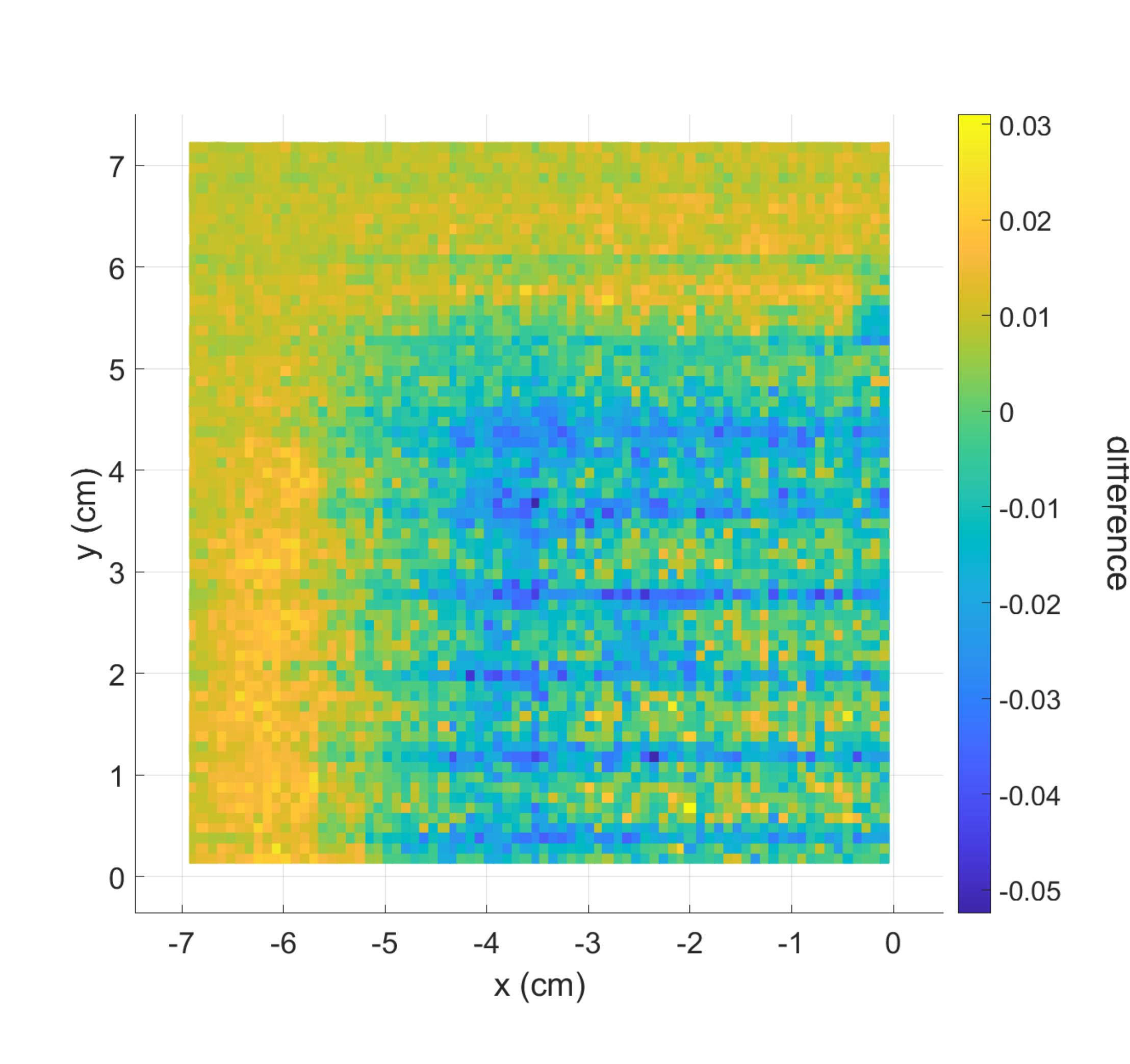}
        \caption{}
    \end{subfigure}
    \caption{a) 2D hit profile of a corner of the GBX Co-60 irradiator beam measured with the silicon pixel detector, and b) difference to a ionization chamber beam profile. The chamber profile between the measurement points was interpolated to get the reference value for the pixel detector results. Note, that due to smaller size of the chamber scan, the values could be compared only on part of the profile.}
    \label{2D_profile}
\end{figure*}

The results from the 2D scan of the corner of the beam are shown in Fig. \ref{2D_profile}. The coordinates for both the pixel detector, and the chamber scans were adjusted by assuming the half value of the profile to be 5 cm from the midpoint of the beam on the x- and y-axis. The pixel detector results are averaged over 6~$\times$~10 pixels (0.9~mm~$\times$~1~mm). As with the 1D-profile, the differences are in the order of a few percent of the profile maximum, although the variance here is higher due to smaller number of counts, and less averaging of pixels. The profile has a structure along the x-axis at the points measured with the edge of the silicon chip that might be a result of charge balance effects at the chip borders. Otherwise, the comparison to the chamber scan shows similar behaviour as the 1D-scan; the pixel detector values at the edge are lower before, and higher beyond the half maximum of the profile. No correction factors, such as $k_1$ and $k_2$, were calculated for the 2D-profile.

%\newpage

\subsection{Repeatability}

\begin{figure*}[!ht]
    \centering
    \includegraphics[width=0.9\textwidth]{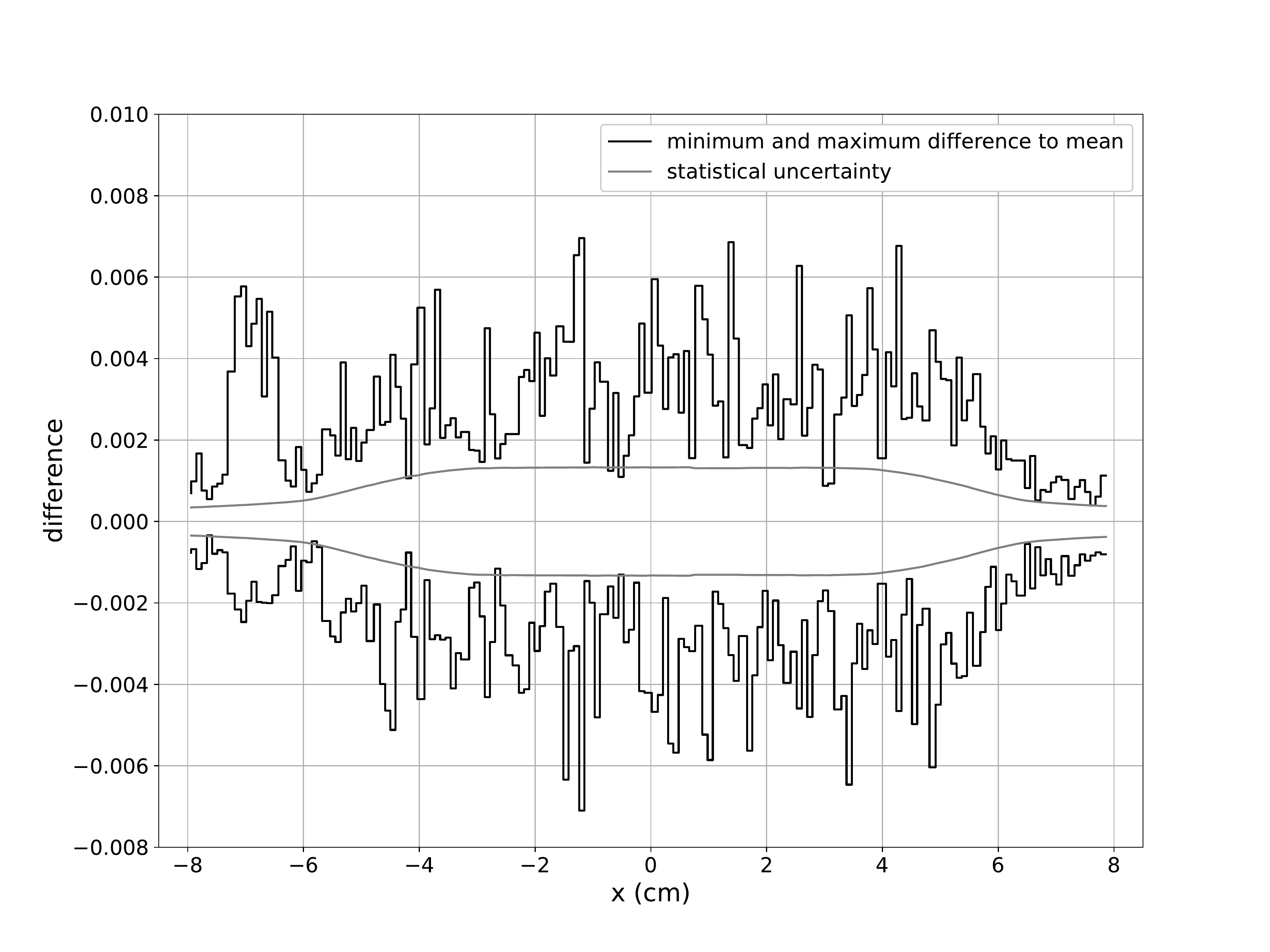}
    \caption{Minimum and maximum difference of the repeated scan results to the mean of all the scans. The profile shape is visualized in Fig. \ref{profiles}a). The scan was repeated six times. The gray line shows the standard deviation calculated from the number of hits in each pixel.}
    \label{repeated}
\end{figure*}

The x-axis scan was repeated multiple times to test the repeatability of the detector. Six profiles were measured during the same day with same detector setup. The detector suffered from errors during some scans, more towards the end of the day. The results from the position of the detector where errors occurred were discarded, leaving gaps in some of the profiles. The data was again averaged over six columns in the x-direction, and the maximum deviations from the mean of all the scans are shown in Fig. \ref{repeated}. For most measurement points, the difference to mean stays below 0.005. The highest deviations occurred at the level part of the beam, where the deviations were caused (in practice) solely by the detector, not the scanner. From \cite{scanner}, a 100~\um\ shift in the detector position would have caused a 0.004 difference in the profile at highest gradients.

\subsection{Count-rate performance}

The performance of the detector in a radiation therapy (RT) photon beam could not be tested within this study, but a crude approximation on the count rate can be made with relative dose rate in the Co-60 and a linac beam. The RT photon beams are pulsed, and a high dose rate is delivered during a single pulse compared to the average dose rate. Usually, the average dose rate is adjusted with the number of pulses in a time interval, and the dose rate during a single pulse cannot be adjusted. The duration of a pulse is in the range of few \si{\micro\second}, and the pause between pulses in the order of milliseconds. From \cite{Beierholm2010}, the dose rate during a single pulse for a Varian 21EX linac with 6~MV voltage at 10~cm depth in water is 60 Gy/s. The dose rate in the Co-60 beam was approximately 6~mGy/s. With an approximation that the dose rate is proportional to flux (which would be the case with an identical electron spectrum), the rate of electrons hitting the detector would be 10000 times higher in a RT photon beam during a pulse. 

From simulations, the number of electrons hitting the detector was approximately $7\cdot10^{-8}$ per source photon. Using the activity of the cobalt source of approximately 100~TBq, and taking into account that, in practice, two photons are emitted per decay, on average 0.35 electrons are hitting the detector in the Co-60 beam in the 25~ns event frame time window. In a radiation therapy beam the same number would be, by using the relative dose rate compared to Co-60 beam, 3500. Considering the total number of pixels (4160), and the readout process, the detector would most likely not be able to measure the beam profile in a radiation-therapy beam due to buffer overflows. However, this is not to say that a similar detector with a smaller pixel size, or different readout electronics would not function sufficiently well in a linac beam. %

\subsection{Spectrum}\label{Spectrum}

\begin{figure*}[!ht] 
    \centering
    \begin{subfigure}[b]{0.48\linewidth}
    \centering
        \includegraphics[height=5.8cm]{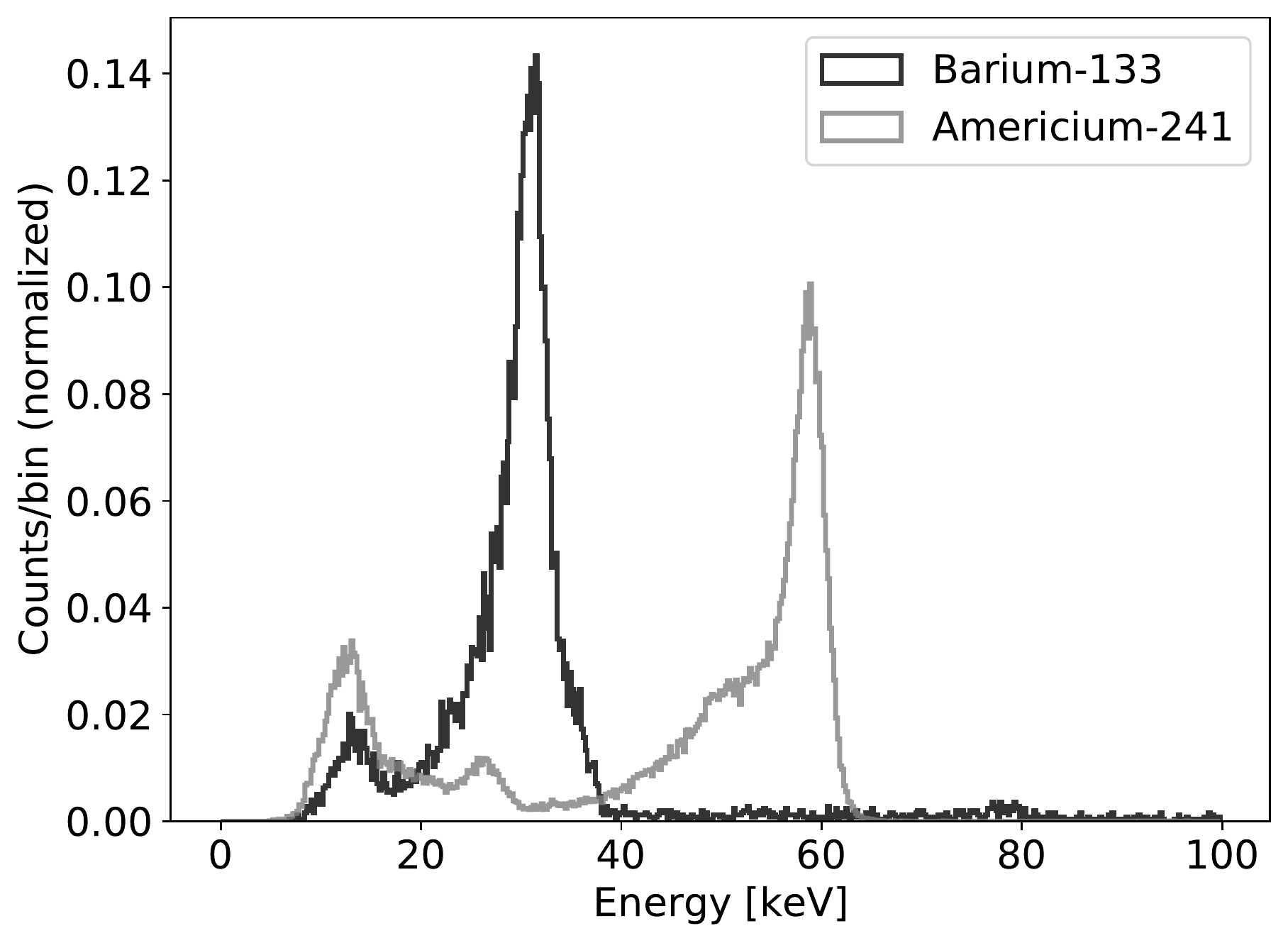}
        \caption{}
    \end{subfigure}
    \begin{subfigure}[b]{0.48\linewidth}
        \centering
        \includegraphics[height=5.8cm]{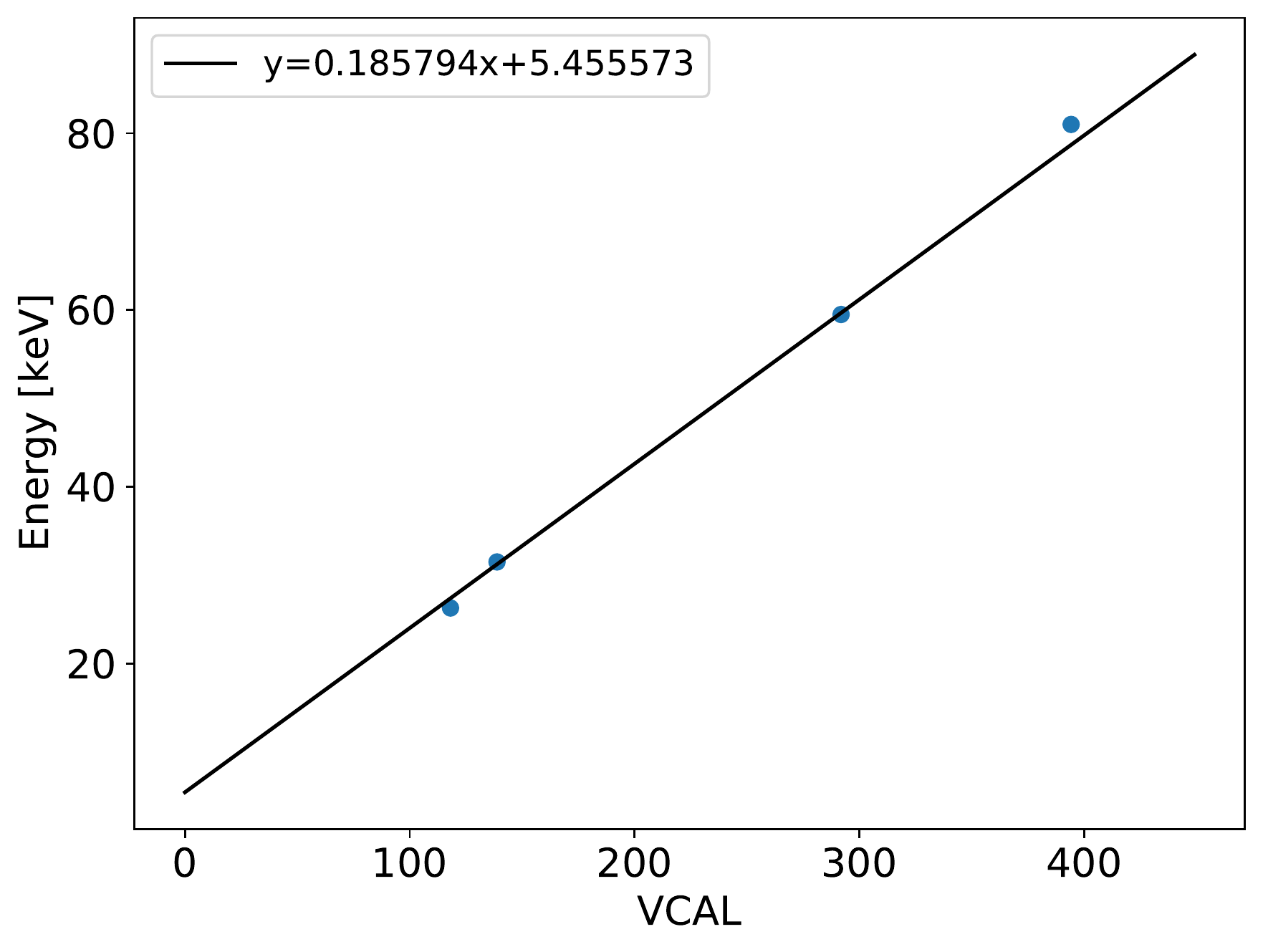}
        \caption{}
    \end{subfigure}
    \caption{(a): Spectra of the Am-241 and Ba-133 radionuclide measurements with the silicon pixel detector. (b): Energy calibration curve obtained from the Am-241 (26.3\,keV, 59.5\,keV) and Ba-133 (30.6\,keV -- 36.0 and 81.0\,keV) peaks.}
    \label{calibration}
\end{figure*}

The detector registers the charges deposited in individual pixels, from which spectra can be constructed. The charges are originally measured with regard to a calibration voltage VCAL and were calibrated to energy values by measuring the spectra of Am-241 and Ba-133 radionuclide sources, and determining the peak centroids of Am-241 gamma (26.3\,keV and 59.5\,keV), and Ba-133 X-ray (30.6\,keV -- 36.0\,keV) and gamma (81.0\,keV) peaks.
Fig.~\ref{calibration} shows the radionuclide spectra taken with the Si pixel detector and the energy calibration curve.

The pulse-height data from a measurement point at the center of the beam was studied more closely. The pixel hits and respective charges are registered and stored in event frames. 
\begin{figure*}[!ht] 
    \centering
    \begin{subfigure}[b]{0.48\linewidth}
    \centering
        \includegraphics[height=5.6cm]{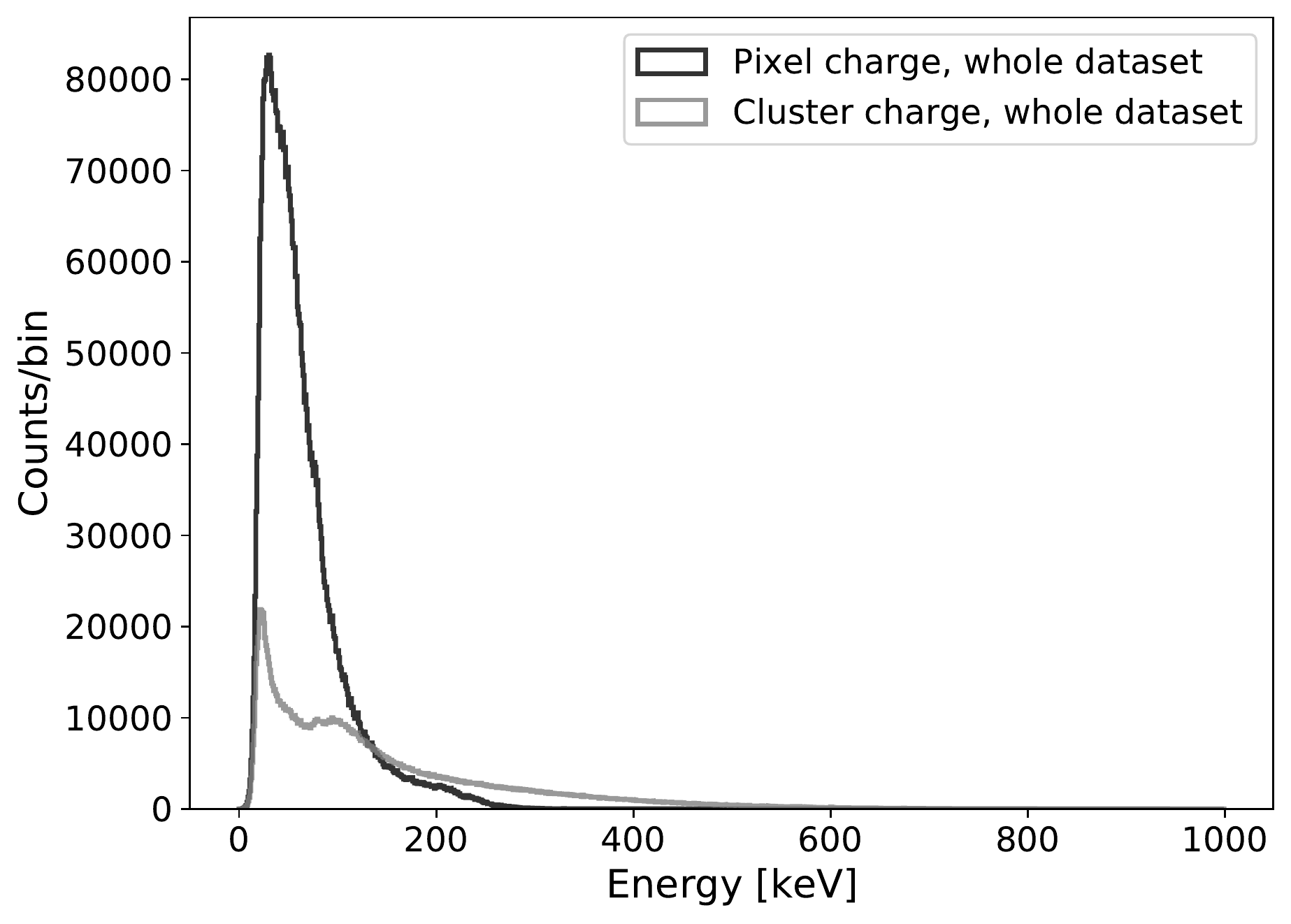}
        \caption{}
    \end{subfigure}
    \begin{subfigure}[b]{0.48\linewidth}
        \centering
        \includegraphics[height=5.6cm]{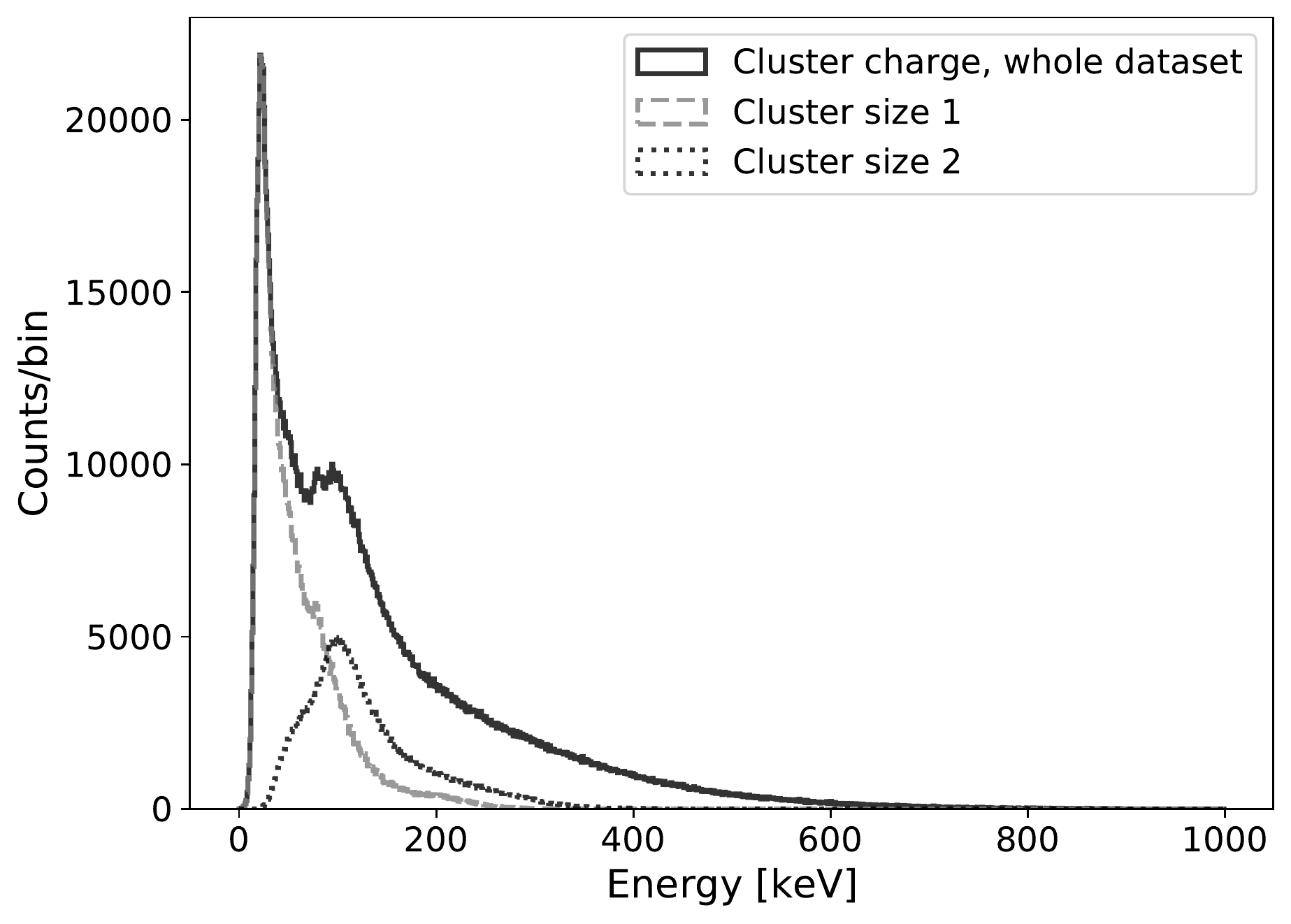}
        \caption{}
    \end{subfigure}
    \caption{Pixel charge and cluster charge. (a): Spectrum of the pixel charge and cluster charge. (b): Cluster charge and the contribution of single hits (cluster size=1) and clusters of size 2.}
    \label{clustercharge}
\end{figure*}
Fig.~\ref{clustercharge}a shows the cumulative spectrum over all detector pixels (pixel charge) and event frames.
Looking at individual event frames, in most of the cases ($36\,\%$ in this data set), single pixels are hit. However, if neighbouring pixels are hit, the combined charge from this cluster of pixels relates usually to energy deposited by a single electron (or positron): In some cases, the charge created by an electron within the volume of a single pixel can diffuse and be registered by the readout of a neighboring pixel (charge-sharing). Also, an electron track can encompass multiple pixels.
\begin{figure*}[!ht] 
    \centering
    \includegraphics[width=0.9\textwidth]{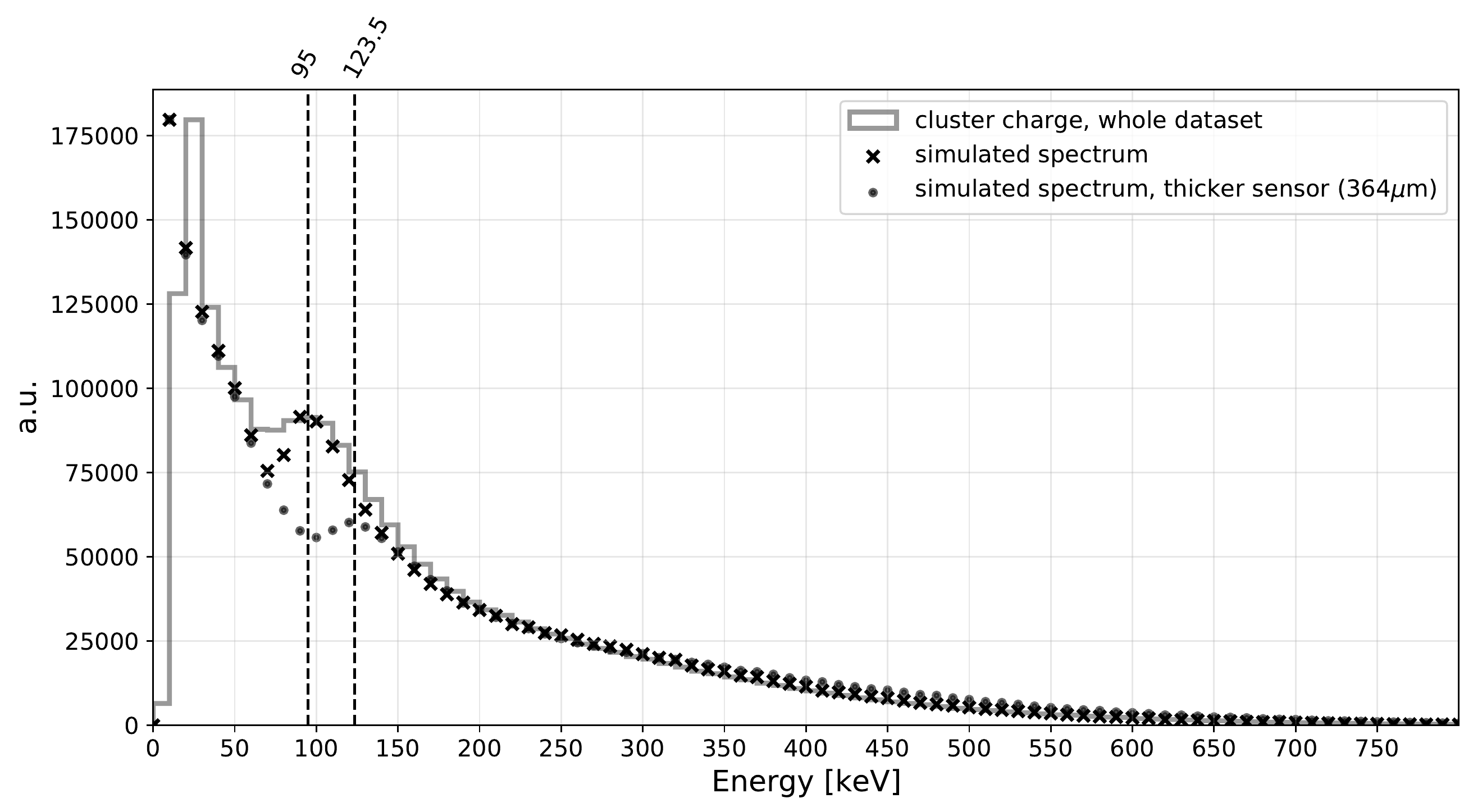}
    \caption{Pulse-height spectra from simulations and from measurements after applying the clustering algorithm. The spectrum was simulated also for a thicker detector to test a hypothesis for the origin of the peak at approximately 95~keV for the original spectrum.
    }
    \label{simspect}
\end{figure*}

To account for the charge-sharing issues, and to measure the full energy deposited by a single electron, a clustering algorithm was developed to sum up the charge of pixels belonging to the same cluster: For every hit pixel, it was checked, whether the nearest neighbours were also hit. This procedure was continued, until no further hit in the vicinity was found (both ends of the track). Charges registered in pixels belonging to the same cluster were then summed, and considered as a single pulse when constructing the pulse height spectrum. The spectrum after applying this clustering algorithm is akin to a spectrum measured using the whole detector as an active volume, but without the effect of summing of charge from multiple electrons hitting the detector within one event.

Fig.~\ref{clustercharge}a shows the spectrum before and after applying the described clustering algorithm. The largest contribution to the cluster charge spectrum are single pixel hits (cluster size 1), followed by clusters with two components. These distributions are depicted in Fig.~\ref{clustercharge}b. The hits in an individual event frame can be visualised as event displays, discussed in section~\ref{eventdisplays}. For example, the event frame depicted in Fig. \ref{events}b contains three separate clusters, one of which contains only a single pixel. 

Because the signal from each pixel is read only if a charge threshold is exceeded in that pixel, regardless of whether it was exceeded in the neighboring pixels, the energy measured for a single cluster is underestimated in cases where a charge below the threshold is collected in one or more pixels along the particle track. A similar effect may occur due to charge sharing. Extrapolation of the energy calibration curve significantly beyond 81~keV may also cause errors in the energy measured by the detector. Pile-up, and counting two electron tracks crossing the same pixel as a single cluster is also possible, but these events happen with low probability. With regard to the detector design of $80\times 52=4160$ pixels, the event with the most hits (51) at this position would relate to an occupancy of only 1.2\%. 

The distribution of the calculated cluster charge aligns quite well with simulation results. This can be seen in Fig.~\ref{simspect}, where the measured spectrum after applying the clustering algorithm is compared to a simulated pulse-height spectrum for the silicon detector.

The measured spectrum in Fig. \ref{clustercharge} has a bump around 95~keV, which falls together with a peak in the simulated spectrum at around the same value. The electron stopping power for silicon is relatively constant in the energy range from 500~keV to 1500~keV, and electrons in this range can travel through the detector chip without significant changes in direction. An 1.25~MeV electron would lose approximately 98~keV along a 280~\um\ track. The dot markers in Fig. \ref{simspect} show the simulated spectrum for a $1.3\times280\,\um=364\,\um$ thick detector. The bump for this spectrum is at approximately 125~keV, and $1.3\times 95$~keV is equal to 123.5~keV. Therefore, the explanation for the bump is most likely electrons passing through the detector in a close to a straight line perpendicular to the detector surface.

\subsection{Event displays}\label{eventdisplays}

\begin{figure*}[!ht]
    \centering
    \includegraphics[width=.9\textwidth]{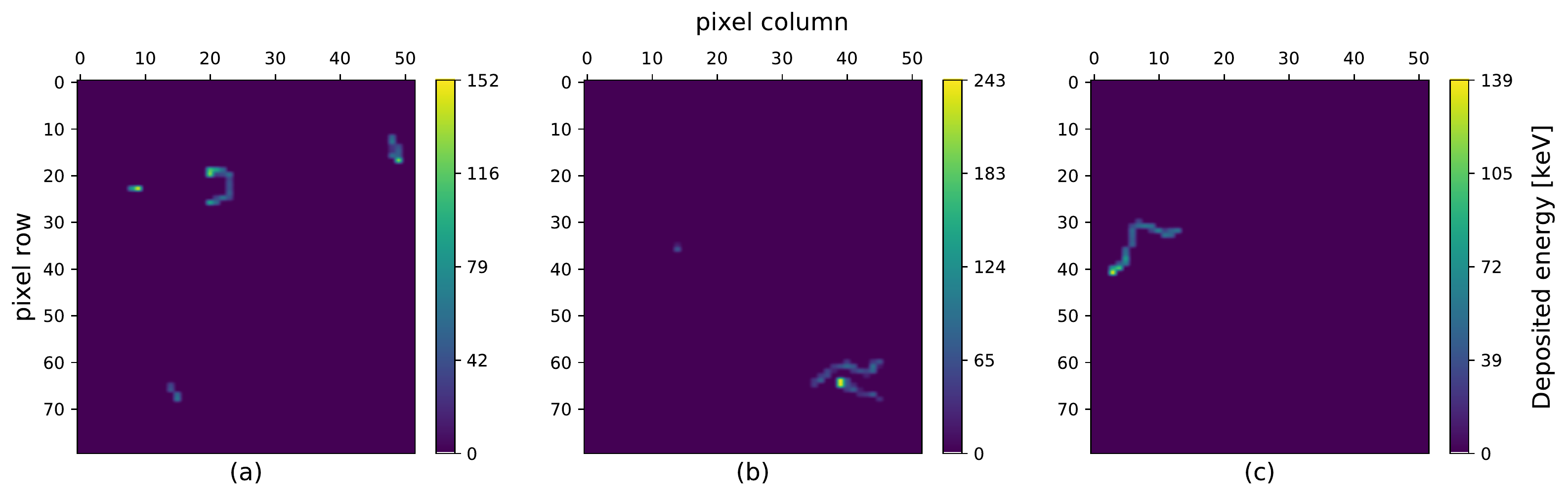}
    \caption{Event displays of three events with a high pixel hit rate, showing several clusters. The color shows the charge deposited in the pixel in keV.
    }
    \label{events}
\end{figure*}
To visualise individual clusters in event frames,
single event frames with a comparatively higher than usual amount of hits were studied more closely and depicted as event displays. We would like to point out here that clusters with more than 10 constituents only contribute to about 1\% of all clusters. Fig.~\ref{events} shows example event displays with clusters of this category. The amount of energy deposited at each end of the electron track hints at the direction of the particle: 
As the electron loses energy along the track, the stopping power increases and more energy will be deposited per unit distance. Also, the electron tracks become more curved at lower energies and electrons can travel longer distances inside a single pixel. Thus, the ending with more energy deposited per pixel is supposedly the ending of the track. However, we only measure the energy imparted in the detector and not the total energy of the electron initially created in a certain distance from the detector chip. Hence, the energy reconstructed, in most cases, is only a fraction of the original electron energy.

%\FloatBarrier

\section{Conclusions}

The silicon pixel detector was successfully used for a measurement of the beam profile in a Co-60 irradiator beam. On an x-axis scan, the results agreed within 0.02 to ionization chamber measurements, and the maximum difference was reduced to approximately 0.01 when correcting for the influence of the detector waterproof casing by Monte Carlo simulations. The average difference was reduced even further by calculating a calibration factor along the scan axis with MC.

We utilized the capability of the pixel detector to record pulse-heights in each pixel, and measured a pulse-height spectrum in the water phantom. Also, we applied an algorithm to analyze charge deposited in a group of neighboring pixels as a single charge pulse. The spectrum constructed like this corresponds approximately to a spectrum scored to the whole active volume of the detector, without the effect of pile-up from multiple electrons hitting the detector in the same event. The results agreed well with a Monte Carlo simulated pulse-height spectrum.

Although the results from the Co-60 beam measurement show promise also for use in a radiation therapy beam, by comparing the dose rates, the detector would most likely not be able to measure a beam profile in a linac photon beam. The situation might be remedied with a smaller pixel size and different electronics.

\section*{Acknowledgements}
This study was performed in the framework of the Academy of Finland project, number 314473, \textit{Multispectral photon-counting for medical imaging and beam characterization}, for which we would like to acknowledge the funding.
%\onecolumn{
%\bibliographystyle{}
\bibliographystyle{elsarticle-num}
\balance
\bibliography{bibliography_pixel_detector_profile_measurements}
%}

\end{document}